\documentclass{jfm}
\usepackage{graphicx,amssymb,amsmath}
\usepackage{natbib,lscape}
\usepackage{subfig}
\usepackage{rotating,framed,color}
\definecolor{shadecolor}{gray}{0.8}

\title[]{Drag cancellation by added-mass pumping}

\author[F. Giorgio-Serchi and G.D. Weymouth ]%
{F. Giorgio-Serchi%
  \thanks{Email address for correspondence: F.Giorgio-Serchi@soton.ac.uk},\ns
G. D. Weymouth}

\affiliation{Southampton Marine and Maritime Institute,\\ Fluid Structure Interaction Group,\\ University of Southampton,\\ Southampton, SO16 7QF, UK\\[\affilskip]
}

\begin{document}

\maketitle

\begin{abstract}
A submerged body subject to a sudden shape-change experiences large forces due to the variation of added-mass energy. While this phenomenon has been studied for single actuation events, application to sustained propulsion requires studying \textit{periodic} shape-change.
We do so in this work by investigating a spring-mass oscillator submerged in quiescent fluid subject to periodic changes in its volume. We develop an analytical model to investigate the relationship between added-mass variation and viscous damping and demonstrate its range of application with fully coupled fluid-solid Navier-Stokes simulations at large Stokes number. 
Our results demonstrate that the recovery of added-mass kinetic energy can be used to completely cancel the viscous damping of the fluid, driving the onset of sustained oscillations with amplitudes as large as four times the average body radius $r_0$. A quasi-linear relationship is found to link the terminal amplitude of the oscillations $X$, to the extent of size change $a$, with $X/a$ peaking at values from 4 to 4.75 depending on the details of the shape-change kinematics. In addition, it is found that pumping in the frequency range of $1-\frac{a}{2r_0}<\omega^2/\omega_n^2<1+\frac{a}{2r_0}$ is required for sustained oscillations.
The results of this analysis shed light on the role of added-mass recovery in the context of shape-changing bodies and biologically-inspired underwater vehicles.
\end{abstract}

\section{Introduction}

The capability of a body to benefit from the added-mass variation induced by changing its shape has generated much interest, particularly by guiding the design of bioinspired propulsion systems. Reports on added-mass variation in self-propelled organisms were first documented in \citet{Daniel1984, Daniel1985} and more recently various contributions have addressed the role of body-shape changes in the unsteady propulsion of aquatic organisms \citep{Kanso2009, Kanso2009b, Candelier2011} based on the early work of \citet{Saffman} and \citet{Lighthill}.

The present work, however, deals with cases where shape change affect the added-mass component in the direction of translation alone, such as when volumetric pulsation or iso-volumetric cross sectional modification take place. This kind of problems, previously accounted for in \cite{Shelly2009}, has implications in the swimming of cephalopods (i.e. squids and octopodes) and the chance to exploit the benefit of their propulsion routine \citep{Trueman1968, Johnson1972} in the frame of soft-bodied underwater vehicles \citep{Weymouth2015b, Giorgio-Serchi2016}.

Earlier works on the dynamics of bodies undergoing single events of rapid volume collapse reported that burst of speed can be achieved by recovering the added-mass kinetic energy of the external flow to augment the momentum flux of the propulsive jet \citep{Weymouth2013, Weymouth2015b}. A goal of the current work is to uncouple added-mass effects from other sources of force and to extend the analysis to periodic shape-changing routines relevant to sustained propulsion. 

To that end, we study a simple spring-mass oscillator subject to prescribed periodic variation of the body volume. By immersing this system in a fluid, the volume change induces added-mass variation, and therefore effects both the effective inertia and damping of the system. This is a novel case of parametric excitation in which `pumping' a parameter of the oscillator is used to induce resonant vibrations \citep{Lavroskii1993, Roura2010}. 
In our system, the shape change is used to completely cancel the substantial fluid drag and drive sustained large amplitude oscillations without any type of propelling jet or background flow. While our results concentrate on spherical bodies, our derivation applies to general isotropic volume change, and the physical principles clearly generalize to any shape change which affects the component of added-mass in the direction of motion.

\section{Governing equations of a volume-changing oscillator}

\begin{figure}
\centering
\includegraphics[trim = 0mm 0mm 0mm -5mm, clip, width=0.5\textwidth]{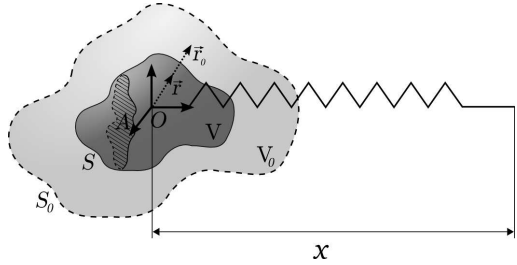}
\caption{Sketch of the volume-changing spring-mass oscillator.}
\label{oscillatorscheme}
\end{figure}

We consider a single degree of freedom oscillator translating in $x$, immersed in a fluid of density $\rho$, figure \ref{oscillatorscheme}. The body is capable of undergoing isotropic changes to its volume $V$ while maintaining its mass $m$ constant. The governing equation is given by
\begin{equation}
m\ddot x+kx = F = -\frac{\partial}{\partial t}\left[C_{11} \rho V \dot x \right] - \frac 12 C_D \rho A \dot x |\dot x|
\label{oscillatordyanmics}
\end{equation}
where $k$ is the spring constant and the fluid forcing $F$ is modelled as the combination of the rate of change of added-momentum as in \citet{Weymouth2013} and a quasi-steady drag force based on the frontal area $A$ as in \citet{Morison1950}. As we consider isotropic volume change, the added-mass coefficient in direction of motion $C_{11}$ and the drag coefficient $C_D$ are idealized as constant. In the remainder of the manuscript we refer to this as the Quasi-Dynamic (QD) forcing model; it being only \textit{quasi}-dynamic as it does not account for any forces that depend on the history of the wake.

Upon expansion and rearrangement, (\ref{oscillatordyanmics}) yields,
\begin{equation}
\left(m+C_{11}\rho V \right) \ddot x + \left(\frac 12 C_D A |\dot x| + C_{11} \dot{V}\right)\rho \dot{x} +k x = 0.
\label{oscillatorundamped}
\end{equation}
By defining the effective inertia and the effective damping coefficients of the system,
\begin{gather}
 m_e = m+C_{11}\rho V \label{me}\\
 c_e = \frac 12 C_D \rho A |\dot x| + C_{11}\rho \dot{V} \label{ce}
\end{gather}
(\ref{oscillatorundamped}) can be rewritten as a classical unforced oscillator
\begin{eqnarray}
m_e \ddot x + c_e \dot x+kx = 0
\label{conciseoscillator}
\end{eqnarray}
Thus, prescribed variation in the volume pumps the oscillator coefficients $m_e$ and $c_e$ and presents the opportunity for parametric excitation and resonance without external forcing. 

The dependence on $\dot{V}$ in (\ref{ce}), is particularly important as it implies that controlled shrinking ($\dot{V}<0$) can be used to cancel the drag for any value of $C_D$, resulting in an undamped oscillator with natural frequency $\omega_n^2 = k/(m+ C_{11}\rho V_0)$ where $V_0$ is the mean volume. We consider two pumping profiles to achieve this: (i) a sharp profile to cancel the drag at all times, (ii) a smooth profile which only zeros the loss on average. 

\subsection{Sharp parametric profile}

We first consider the sharp profile. As our goal is an undamped oscillator, we assume an oscillation of the form
\begin{eqnarray}
 x = X\cos(\omega t)
 \label{assumed_x}
\end{eqnarray}
where $X$ is the amplitude of displacement of the oscillator, figure \ref{oscillatorscheme}. 

A radius profile for drag cancellation is readily derived by defining the body surface $S$ and a radius vector $\vec{r}$ from the origin to $S$. Since isotropic shape changes ensure $\dot{V} = \dot{r} S$, (\ref{ce}) gives $c_e=0$ when
\begin{eqnarray}
\dot{r}=-\frac 12 \frac{C_D}{C_{11}}\frac{A}{S}|\dot{x}|
\label{radius}
\end{eqnarray}
The proportionality to $-|\dot{x}|$ in (\ref{radius}) implies that the volume must continually shrink. To avoid this while maintaining zero damping we apply a step increase in $r(t)$ back to its maximum when $\dot x=0$, i.e twice per oscillation. This can be conveniently expressed by defining a modulo temporal function $t^*$
\begin{eqnarray}
t^* = t \bmod \frac \pi\omega
\end{eqnarray}
allowing us to define the radius profile as
\begin{eqnarray}\label{sharp}
 r(t^*) = r_0+a\cos(\omega t^*)
\end{eqnarray}
where $r_0$ is the norm of the original $\vec{r}$ and $a$ is the amplitude of oscillation. Due to the discontinuity, we call this the `sharp' profile, shown in figure \ref{comparison3}.

The variation of $r$ will also change the effective mass proportional to $\left(r/{r_0}\right)^3$
\begin{eqnarray}\label{me2}
 m_e = m+C_{11}\rho V_0 \left(1+3\frac a{r_0}\cos(\omega t^*)+\mathcal{O}\left(\frac{a^2}{r^2_0}\right)\right)
\end{eqnarray}
Assuming the damping has been cancelled, dividing the unforced oscillator equation through by $m_e$ recovers the classic parametric oscillator system
\begin{gather}
\ddot x + \Omega^2 x = 0 \\
\Omega^2 = \omega_n^2\left(1-3\frac{C_{11}}{m^*+C_{11}}\frac a{r_0}\cos(\omega t^*)\right)\label{Omega}
\end{gather}
where the mass ratio is $m^*=m/(\rho V_0)$ and we have neglected nonlinear terms in $a/r_0$. Pumping a parametric oscillator with period approximately half the natural period leads to parametric resonance and an exponential increase in oscillation amplitude \citep{Lavroskii1993}. As the period of the sharp waveform (\ref{sharp}) is $\pi/\omega$ due to the modulo in the definition of $t^*$, this requires $\omega\approx\omega_n$. The window for parametric resonance is $\omega^2/\omega_n^2 = 1 \pm f/2$ where $f=3\frac{C_{11}}{m^*+C_{11}}\frac a{r_0}$ is the parametric amplitude from (\ref{Omega}).

When pumping within this resonant window, any initial perturbation of the system in $x$ will increase in magnitude until there is a net loss of power due to the fluid forces over a cycle,  $\Delta P=\int_0^T F\dot x dt$. Note that the total fluid force must be considered, not just the damping, because the added-inertia contributes to the power balance. Substituting the Quasi-Dynamic model for $F$ and setting $\Delta P=0$ gives
\begin{eqnarray}
-\frac 12 C_D \rho \frac {A_0}{r_0^2}\int_0^T r^2 |\dot x|\dot x^2 dt-C_{11}\rho \frac {S_0}{r_0^2}\int_0^T\left(\frac 13r^3\ddot x+r^2 \dot r \dot x\right)\dot x dt = 0
\end{eqnarray}
where we have used the fact that areas scale proportional to $(r/r_0)^2$ and $V_0=\frac{1}{3}S_0 r_0$. Substituting the waveforms for $x$ and $r$ (\ref{assumed_x},~\ref{sharp}) we find that all the nonlinear $a/r_0$ terms cancel and we obtain a simple linear relationship between the amplitude of pumping and the oscillation response
\begin{eqnarray}\label{response}
X = a \frac{C_{11}}{C_D}\frac{S}{A}
\end{eqnarray}
Plugging this relationship into the effective damping (\ref{ce}) we see that we have in fact \textit{over-compensated} for the drag; the body experiences thrust throughout the cycle. This thrust is required to compensate for the power spent on pumping the added-inertia.

\subsection{Smooth pumping profile}
\begin{figure}
\centering
\includegraphics[trim = 0mm 0mm 0mm 0mm, clip, width=0.9\textwidth]{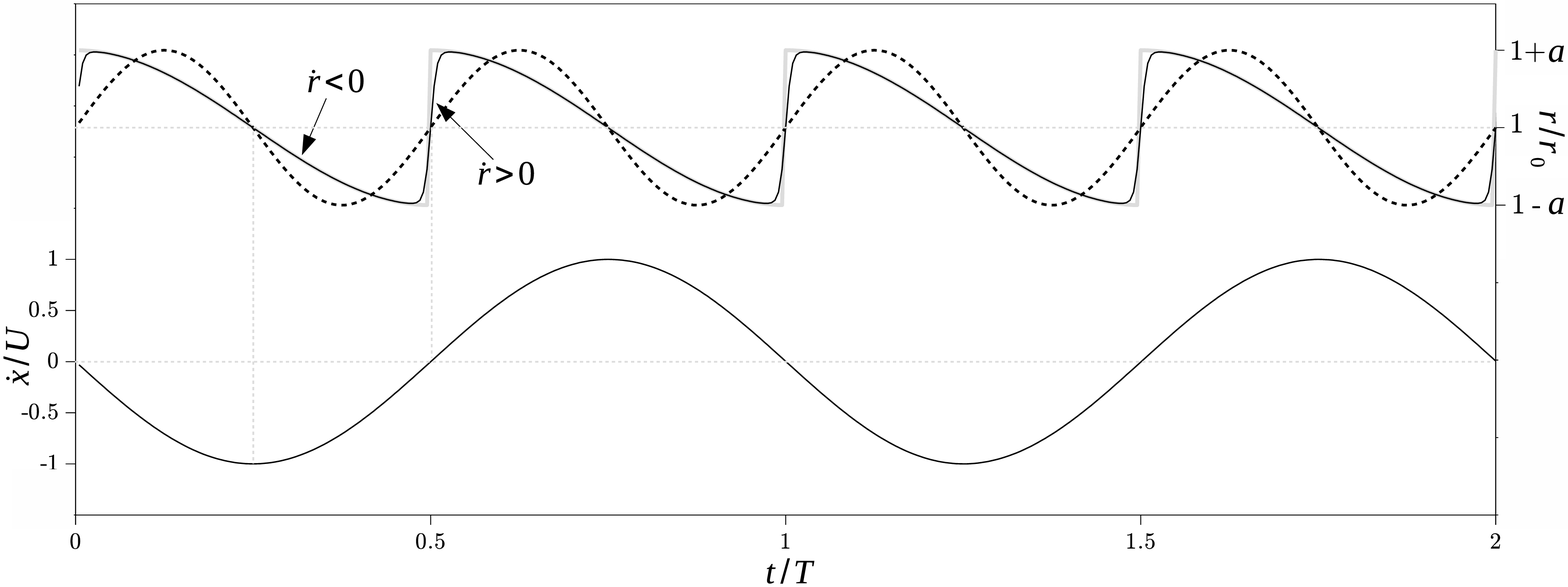}
\caption {Upper part of the plot: profile of radius variation for a spherical oscillator pulsating around the average radius $r_0$ with sharp profile (\ref{sharp}), gray, continuous sharp profile (\ref{tilde}), black, and smooth profile (\ref{smooth}), dashed black. In the lower part of the plot: initially assumed velocity profile (\ref{assumed_x}) with $U=\omega X$ being a reference velocity.}
\label{comparison3}
\end{figure}

The previous results generalize to cases without discontinuous shape change. Consider instead a smooth profile that still has period $\pi/\omega$ and amplitude $a$
\begin{eqnarray}\label{smooth}
r(t) = r_0+a\sin(2 \omega t)
\end{eqnarray} 
As before, the variation in $r$ leads to variation in $m_e$ and parametric resonance when pumping with $\omega\approx\omega_n$. Using \ref{smooth} and neglecting nonlinear terms in $a/r_0$ we find a new relationship between $X$ and $a$ for which $\Delta P=0$
\begin{eqnarray}\label{smooth_reponse}
X = \frac{3\pi}8 a \frac{C_{11}}{C_D}\frac{S}{A}
\end{eqnarray}
Comparing to (\ref{response}), we see the smooth profile increases the response for a given amplitude because the rate of shrinking is double that of the sharp profile, figure~\ref{comparison3}, but the irregular cancellation of the drag over a cycle limits the increase to 18\%.

\section{Results for a spherical, volume-changing oscillator}

We next present the nonlinear response of the volume-changing oscillator using the Quasi-Dynamic forcing model and fully coupled Navier-Stokes simulations. We restrict our analysis to a spherical body with mass ratio $m^*=1$ for brevity. 

\subsection{Results for the Quasi-Dynamic model}

\begin{figure}
\centering
\includegraphics[trim = 0mm 0mm 0mm 0mm, clip, width=0.95\textwidth]{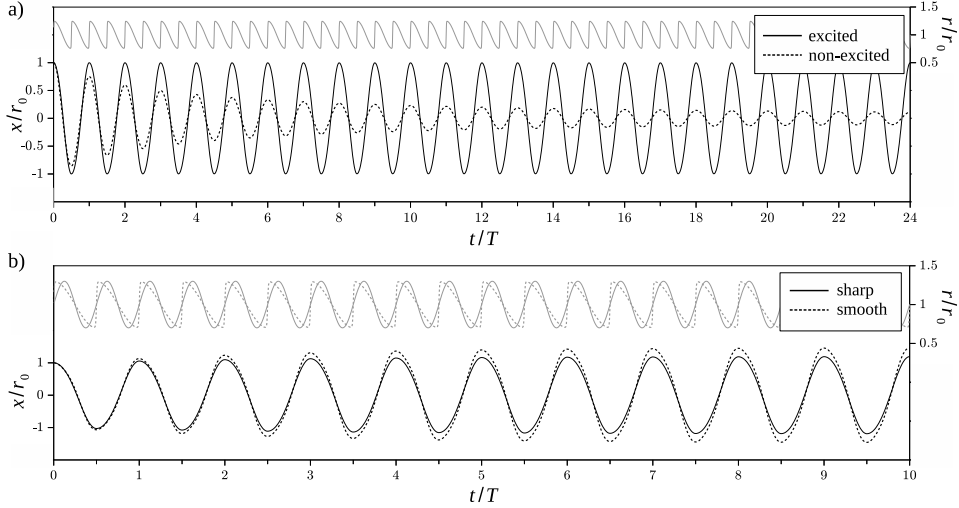}
\caption{Comparison of radius variation (upper part) and resulting oscillations (lower part) for: (a) a non-excited spherical oscillator and an oscillator pumped using the sharp profile with $a=0.225r_0$ at the natural frequency, (b) oscillators excited by the sharp and smooth profiles with the same frequency and amplitude as (a).}
\label{zerodamping1}
\end{figure}

To apply the Quasi-Dynamic forcing model we use the standard potential flow added-mass for a sphere $C_{11}=\frac 12$ and the reference drag coefficient  for oscillating spheres at intermediate Reynolds number $C_D=0.45$ \citep{Mei1994}. With these values set, (\ref{conciseoscillator}) is integrated numerically using a standard high-order method with adaptive time-stepping. Figure~\ref{zerodamping1}(\textit{a}) compares the response of the oscillator when not excited to the case when it is excited using the sharp pumping profile (\ref{sharp}) with $a/r_0=0.225$ and $\omega=\omega_n$. As expected, the un-excited oscillator exhibits the classic `under-damped' response. In contrast the pumped oscillator exhibits sustained oscillations with amplitude $X=r_0$, in agreement with (\ref{response}), demonstrating complete drag cancellation.

Figure \ref{zerodamping1}(\textit{b}) compares the oscillator response between the sharp and smooth pumping profiles using the same $a$ and $\omega$. The `smooth' sinusoidal profile exhibits slightly increased terminal amplitudes, in agreement with (\ref{smooth_reponse}). The response also contains more evidence of higher harmonics than the `sharp' response due to the application of positive and negative damping at different points in the cycle.

\begin{figure}
\centering
\includegraphics[trim = 0mm 0mm 0mm 0mm, clip, width=1\textwidth]{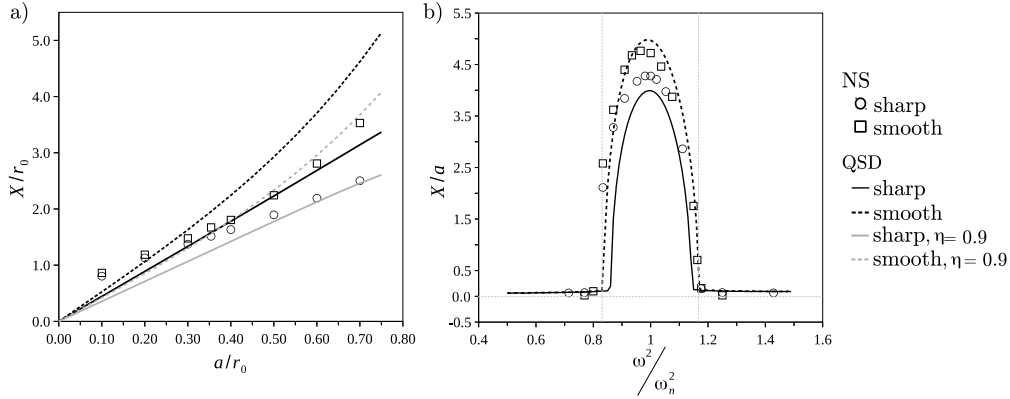}
\caption{(a) Oscillator response to varied pumping amplitude. The symbols are the Navier-Stokes simulations, the lines are the Quasi-Dynamic (QD) model results. The gray $\eta=0.9$ lines are for the QD model with 90\% added-mass recovery (\ref{ce_eta}). (b) System response to varied pumping frequency. The gray vertical lines mark the $\omega^2/{\omega_{n}}^{2} = 1 \pm f/2$ envelop within which parametric resonance is expected (\ref{Omega}).}
\label{phaseAndAmplitude}
\end{figure}

We extend the comparison between the sharp and smooth profiles across the amplitude range of $0 \le a/r_0\le 0.8$, pumping at the natural frequency, figure \ref{phaseAndAmplitude}(\textit{a}). Each case is integrated until the oscillation amplitudes attain steady values. Finally, we study the effect of out-of-phase parametric forcing in figure \ref{phaseAndAmplitude}(\textit{b}) by pumping across the frequency range of $0.5\le\omega^2/{\omega_{n}}^{2}\le1.5$ at $a/r_0=0.35$.

\subsection{\label{FSI}Results for fully coupled Navier-Stokes simulations}

The Quasi-Dynamics forcing model neglects wake history effects, must be supplied with values for the added-mass and drag coefficients, and assumes that all of the added-momentum of the fluid is recoverable via shape-change. A set of three-dimensional fully coupled Navier-Stokes simulations are used to asses the limitations of the model, and verify the ability of added-mass variation to cancel drag forces indefinitely. 

To ensure relevance to high-speed biological and biomimetic systems, we use a large Stokes number, $\varepsilon = \sqrt{\frac{r_0^2\omega}{2\nu}} = 25$. The Reynolds number based on the maximum speed $U=\omega X$ and diameter $D=2r_0$ is then $Re = 4\frac X{r_0}\varepsilon^2$ which depends on the response of the system but will always be less than $10^4$. \citet{Weymouth2015b} introduced a shape-change rate to characterize the ability of a body to recapture added-mass energy $\sigma^*=\frac{\dot V}{A U}\sqrt{Re}$.
Added-mass recovery was estimated to require $\sigma^*>9$ while experiments performed at $\sigma^*= 77$ were found to provide a measurable contribution of added-mass recovery on thrust. Applying this factor to the spherical oscillator gives $\sigma^* = 4\dot r\sqrt{Re}/(\omega X)$ which is again dependant on the magnitude of response, but will be less than $100$.

The simulations use the Boundary-Data-Immersion-Method (BDIM) \citep{Weymouth2011, Weymouth2015}, a robust immersed boundary method suitable for dynamic Fluid-Structure-Interaction (FSI) problems. This method solves both the Navier-Stokes equations and the dynamics of the volume-changing sphere as it travels through the fluid. Previous work has validated this approach for a variety of FSI problems including a deforming self-propelled model of a fast escaping octopus \citep{Weymouth2013}.

The computational domain extends $8r_0$ from the maximum excursion of the sphere in every direction. No-slip and no-penetration boundary conditions are imposed on the solid-fluid interface. Pressure outlet conditions are defined at the domain boundaries to enable fluid mass flux to compensate for the volume changes of the body. The fluid equations are solved using a second-order finite-volume method in space and an explicit second-order method in time. The solid equation is simply (\ref{oscillatordyanmics}) with the fluid force $F$ calculated from the fluid simulation. All simulations are performed with 64 points per diameter in the region in which the sphere travels, with exponential grid stretching applied in the far-field. The choice for this degree of resolution is based on the grid convergence analysis reported in table \ref{grid_convergence}.

\begin{table}
  \begin{center}
\def~{\hphantom{0}}
  \begin{tabular}{lcccc}
      $2r_0/h$  &  32 & 45 & 64 & 96 \\
      $X/a$ & 4.098 &  4.328 &  4.723 &  4.888 \\
      $\% error$ &    0.163 &    0.115 &    0.034 &    -- \\
  \end{tabular}
  \caption{Grid spacing $h$ convergence results using the smooth pumping profile with $a/r_0=0.35$ and $\omega=\omega_n$. The percent error is computed relative to the finest grid.}
  \label{grid_convergence}
  \end{center}
\end{table}

\begin{figure}
\centering
\includegraphics[trim = 0mm 0mm 0mm 0mm, clip, width=1\textwidth]{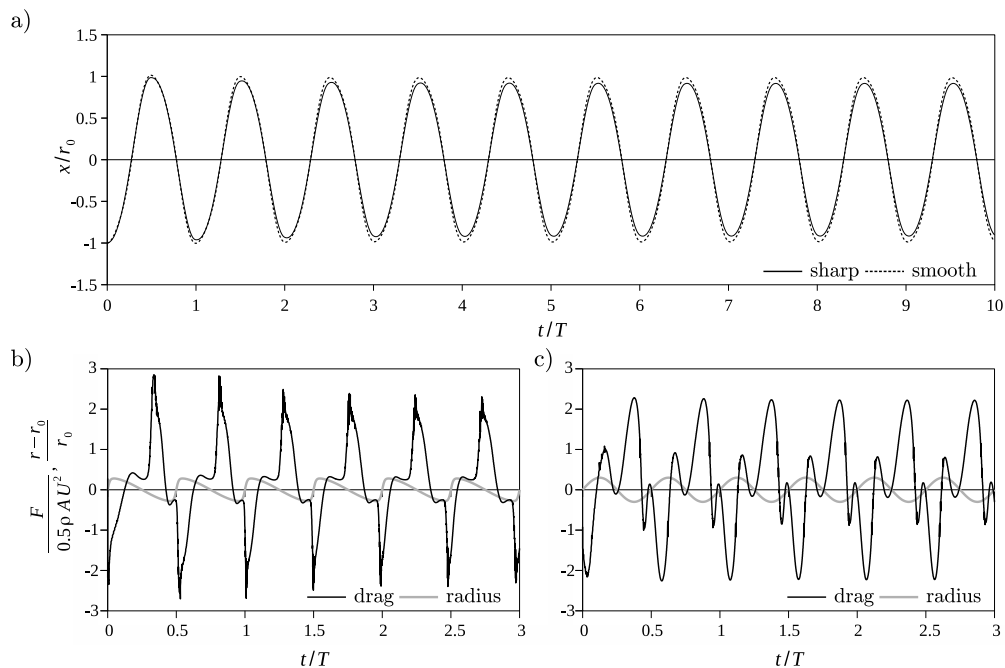}
\caption{Results from Navier-Stokes solution of the dynamics of the shape-changing oscillator: (a) amplitude of the oscillation for the sharp and smooth pumping; (b) and (c) time-varying drag coefficient of the oscillator respectively for the sharp (b) and smooth (c) pumping.}
\label{zerodamping2}
\end{figure}

\begin{figure}
	\centering
	\begin{minipage}{0.46\textwidth}\centering
	\begin{snugshade}
		\scalebox{0.8}{Sharp} 
	\end{snugshade}
	\end{minipage}
	\hspace{2mm}
	\begin{minipage}{0.46\textwidth}\centering
	\begin{snugshade}
		\scalebox{0.8}{Smooth}
	\end{snugshade}
	\end{minipage}
	\begin{minipage}{5mm}
	$\ $
	\end{minipage}\\
	\begin{minipage}{0.47\textwidth}
		\includegraphics[trim=12cm 8.5cm 12cm 8.5cm, clip=true, 
		width=\textwidth]{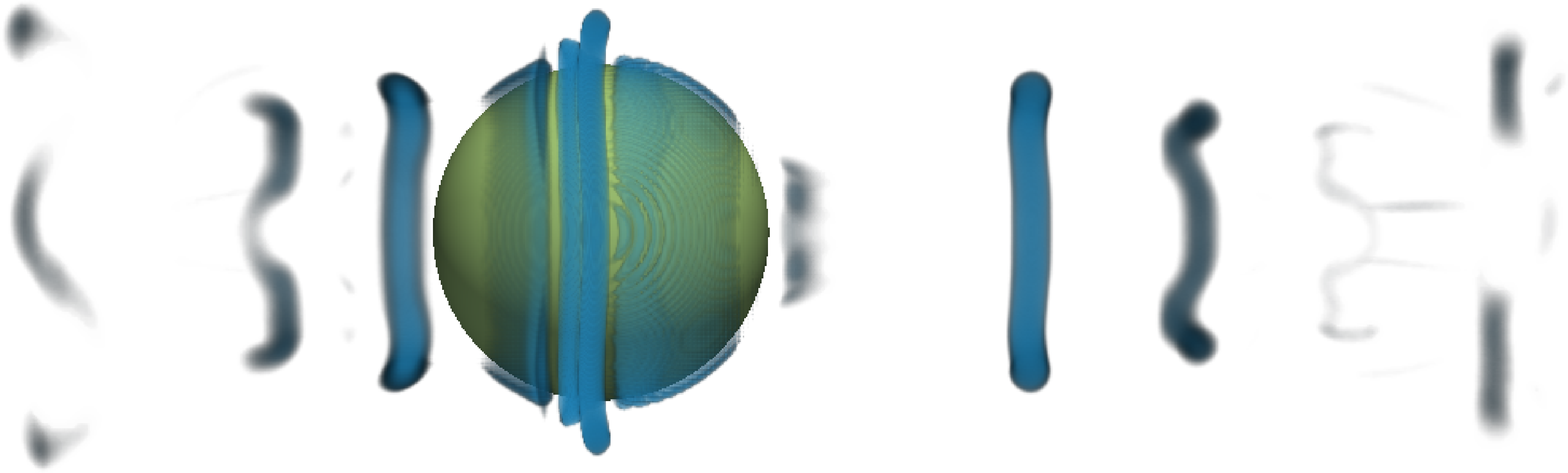}
	\end{minipage}
	\begin{minipage}{0.47\textwidth}
		\includegraphics[trim=12cm 8.5cm 12cm 8.5cm, clip=true, 
		width=\textwidth]{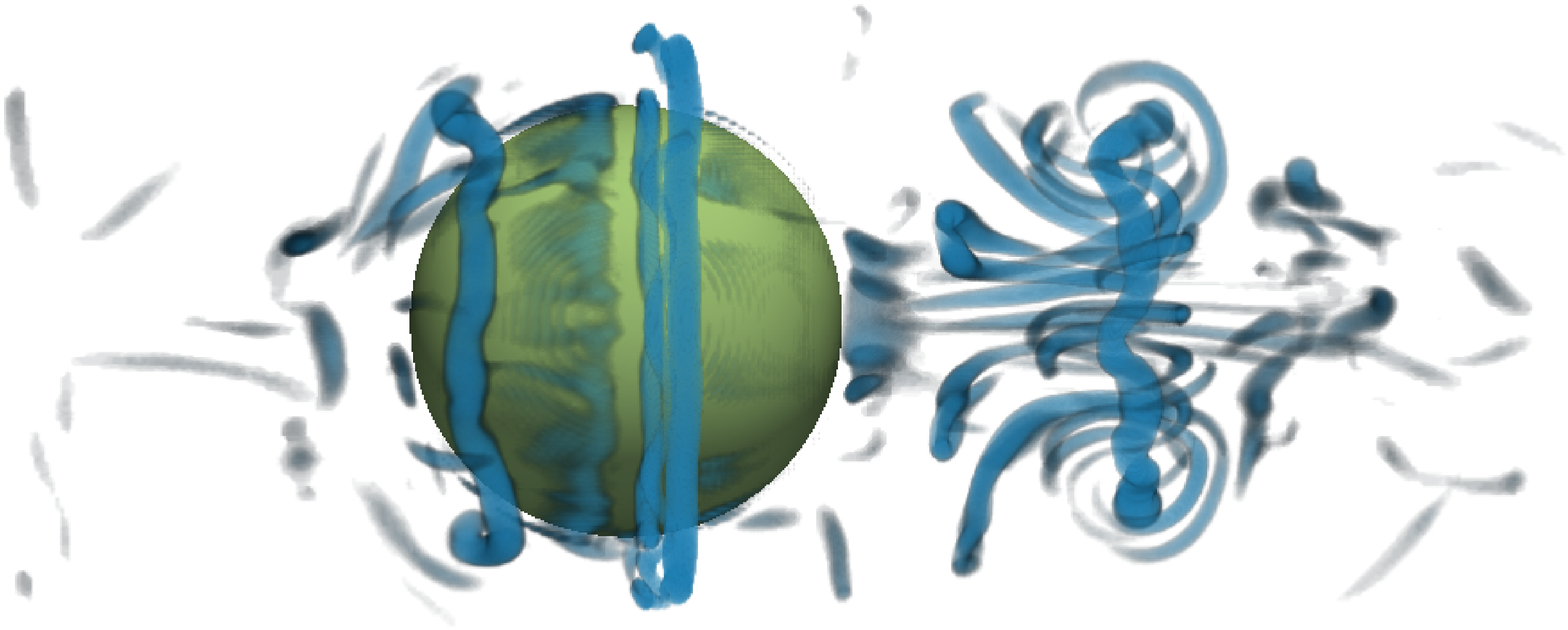}
	\end{minipage}
	\rotatebox[origin=c]{270}{
	\begin{minipage}{24mm}\centering
	\begin{snugshade}
		\scalebox{0.8}{$t/T = 0.2 $} 
	\end{snugshade}
	\end{minipage}}\\
	\begin{minipage}{0.47\textwidth}
		\includegraphics[trim=12cm 8.5cm 12cm 8.5cm, clip=true, 
		width=\textwidth]{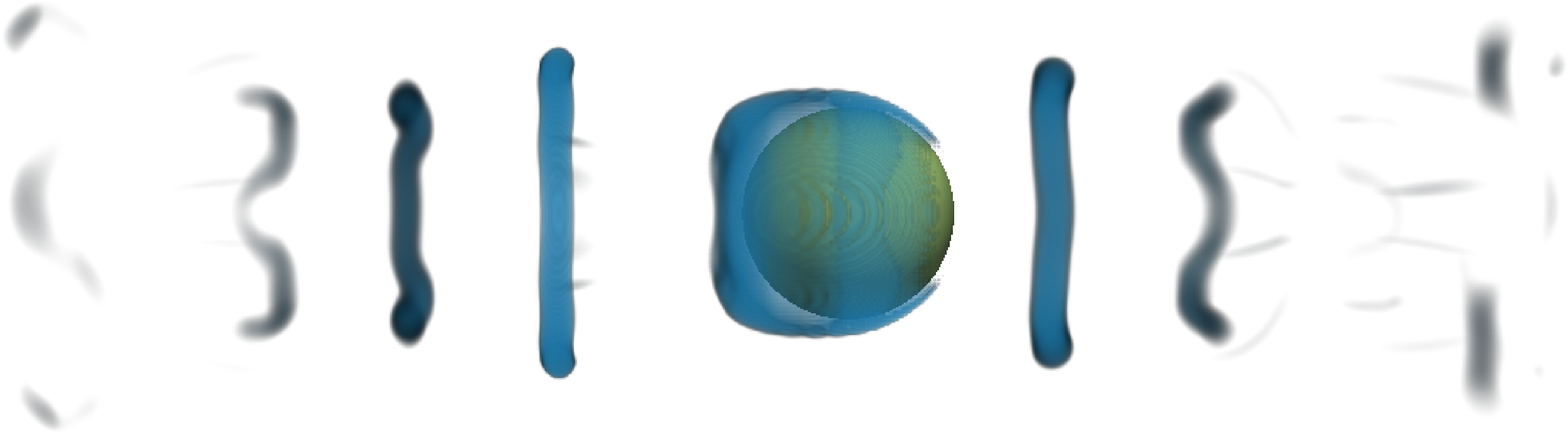}
	\end{minipage}
	\begin{minipage}{0.47\textwidth}
		\includegraphics[trim=12cm 8.5cm 12cm 8.5cm, clip=true, 
		width=\textwidth]{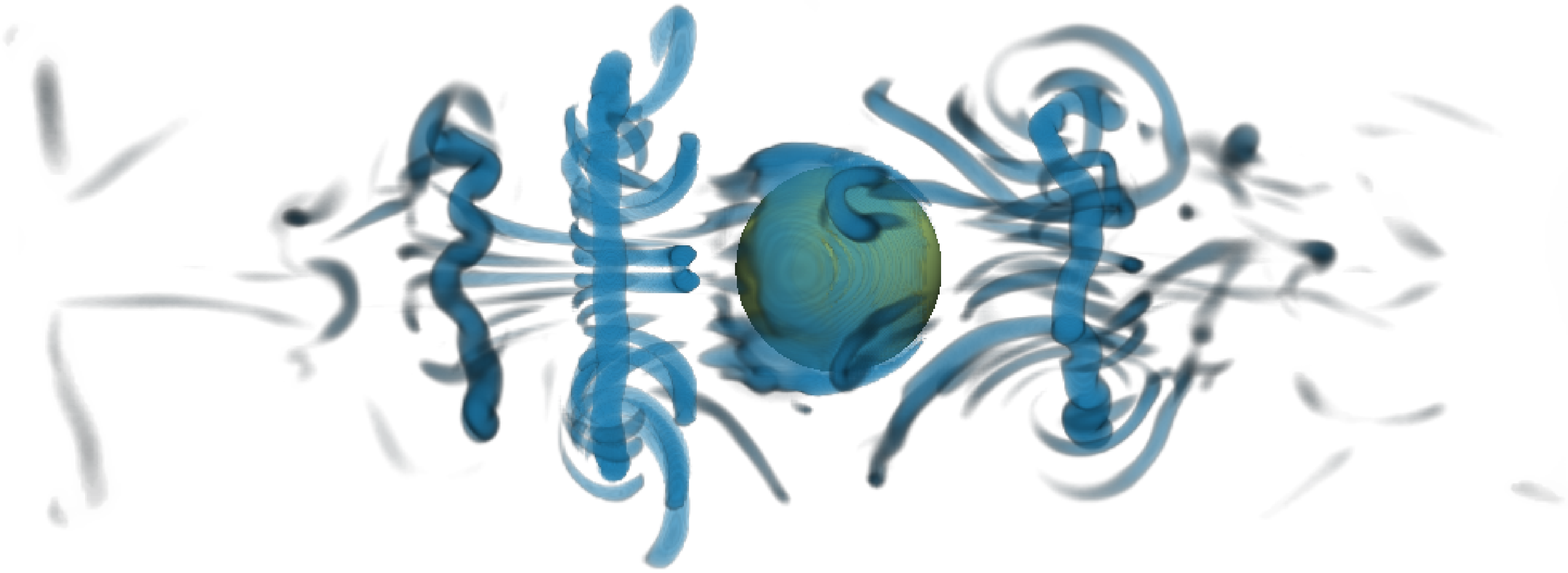}
	\end{minipage}
	\rotatebox[origin=c]{270}{
	\begin{minipage}{24mm}\centering
	\begin{snugshade}
		\scalebox{0.8}{$t/T = 0.4 $} 
	\end{snugshade}
	\end{minipage}}\\
	\begin{minipage}{0.47\textwidth}
		\includegraphics[trim=12cm 8.5cm 12cm 8.5cm, clip=true, 
		width=\textwidth]{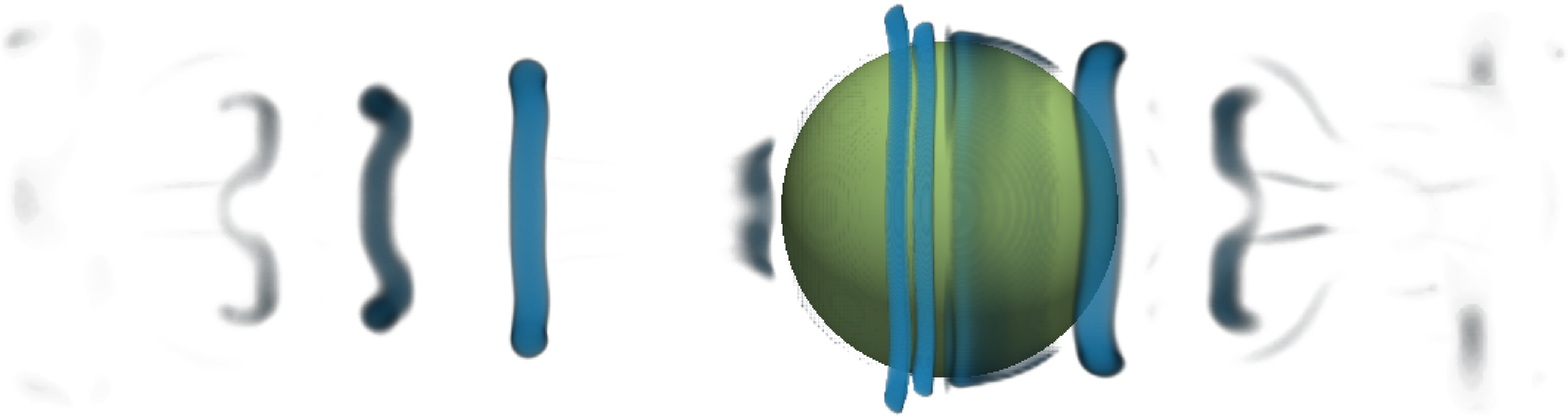}
	\end{minipage}
	\begin{minipage}{0.47\textwidth}
		\includegraphics[trim=12cm 8.5cm 12cm 8.5cm, clip=true, 
		width=\textwidth]{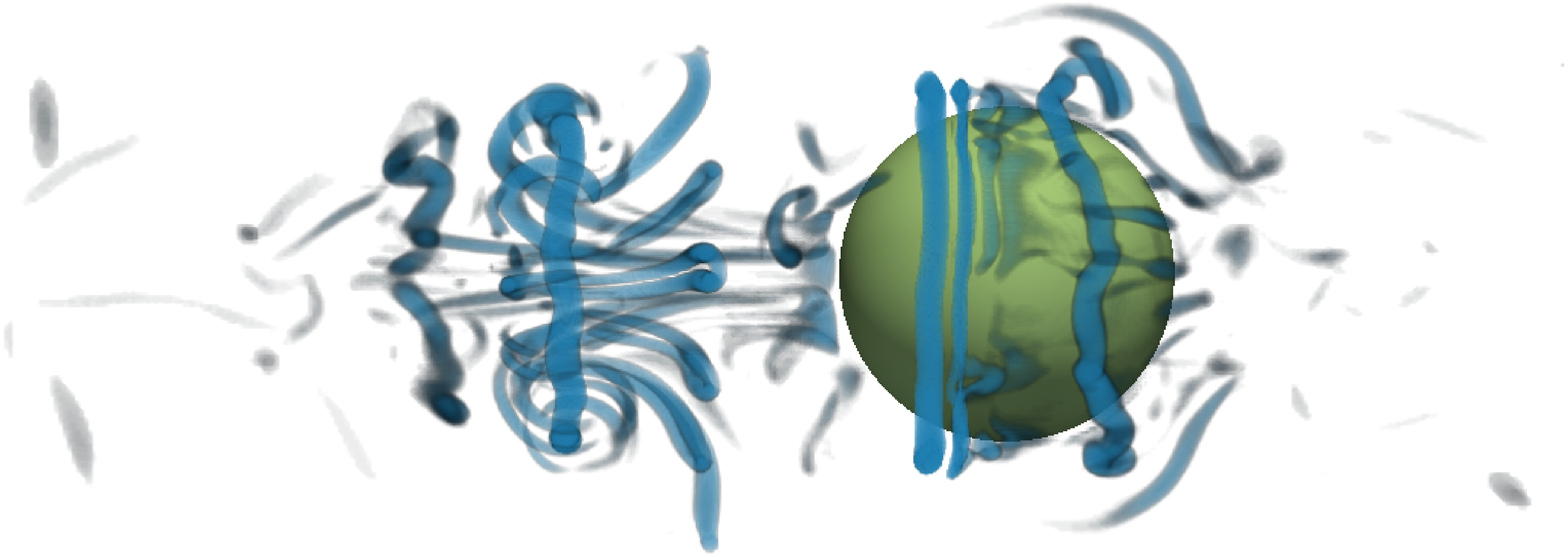}
	\end{minipage}
	\rotatebox[origin=c]{270}{
	\begin{minipage}{24mm}\centering
	\begin{snugshade}
		\scalebox{0.8}{$t/T = 0.6 $} 
	\end{snugshade}
	\end{minipage}}\\
	\begin{minipage}{0.47\textwidth}
		\includegraphics[trim=12cm 8.5cm 12cm 8.5cm, clip=true, 
		width=\textwidth]{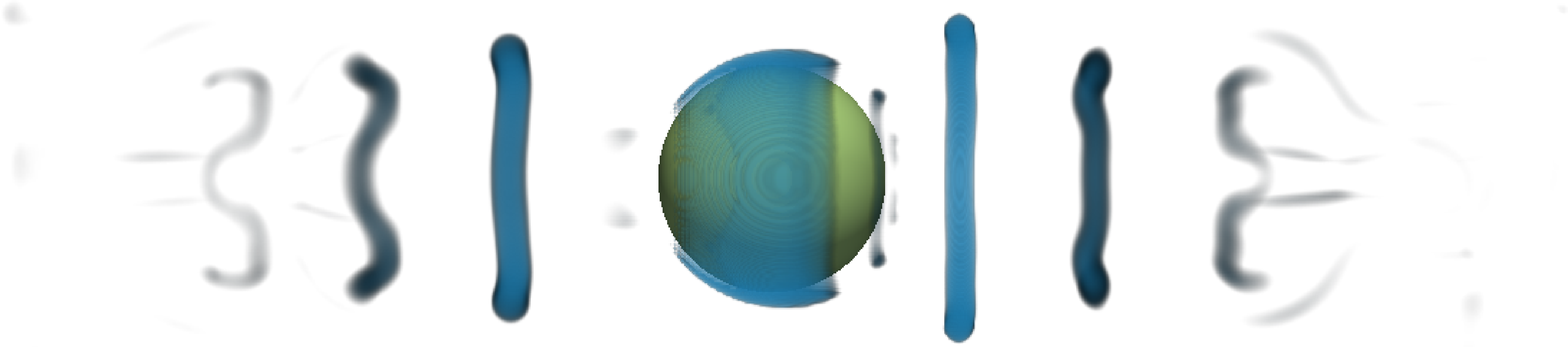}
	\end{minipage}
	\begin{minipage}{0.47\textwidth}
		\includegraphics[trim=12cm 8.5cm 12cm 8.5cm, clip=true, 
		width=\textwidth]{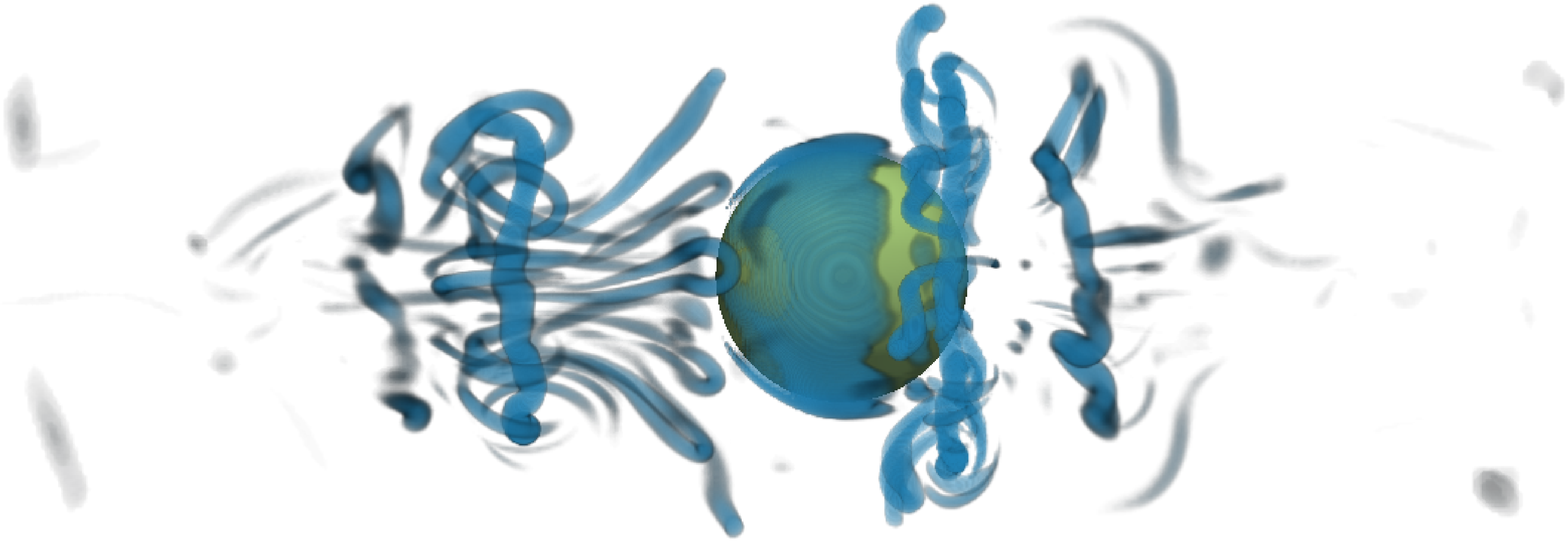}
	\end{minipage}
	\rotatebox[origin=c]{270}{
	\begin{minipage}{24mm}\centering
	\begin{snugshade}
		\scalebox{0.8}{$t/T = 0.8 $} 
	\end{snugshade}
	\end{minipage}}\\
	\begin{minipage}{0.47\textwidth}
		\includegraphics[trim=12cm 8.5cm 12cm 8.5cm, clip=true, 
		width=\textwidth]{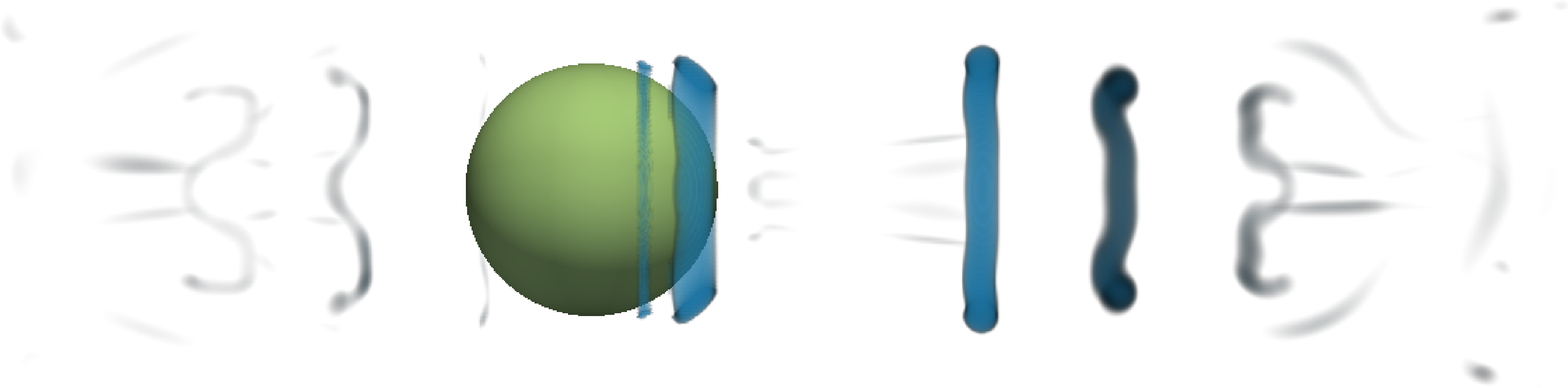}
	\end{minipage}
	\begin{minipage}{0.47\textwidth}
		\includegraphics[trim=12cm 8.5cm 12cm 8.5cm, clip=true, 
		width=\textwidth]{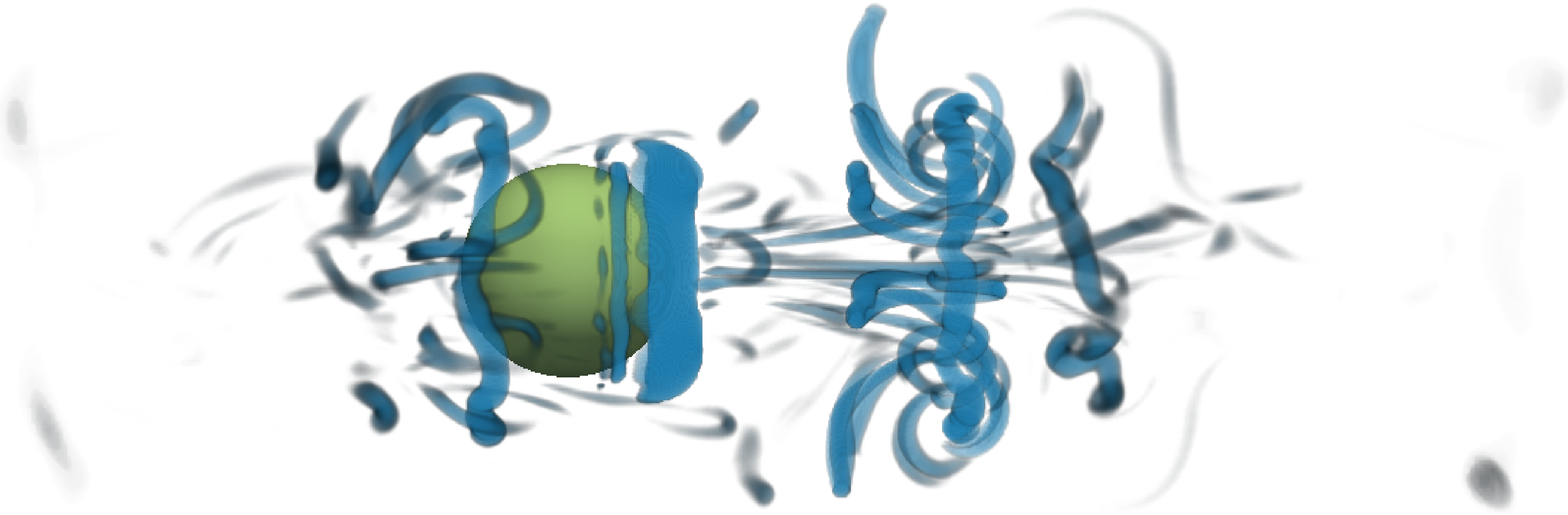}
	\end{minipage}
	\rotatebox[origin=c]{270}{
	\begin{minipage}{24mm}\centering
	\begin{snugshade}
		\scalebox{0.8}{$t/T = 1.0 $} 
	\end{snugshade}
	\end{minipage}}
	\caption{Evolution of the $\lambda_2$ vortex criterion during one oscillation after attainment of zero-damping regime in response to the sharp (left) and smooth (right) radius variation profile at five moments along the cycle. See movie 1 in the supplementary material.}
\label{lambda2}
\end{figure}

The last methodology concern is that the discontinuity in the sharp pumping profile would induce infinite forces in the fluid domain. In order to avoid this, a continuous version is formulated by parametrizing the rate of inflation, $\Delta$, such that
\begin{eqnarray}\label{tilde}
 \tilde t = t^*+T\frac{\tanh(\Delta (1/2-t^*/T))-\tanh(\Delta (t^*/T))}4
\end{eqnarray}
The result is continuous, but as $\Delta\rightarrow\infty$ we recover the modulo function. We choose $\Delta=40$ such that when $\tilde{t}$ is used in (\ref{sharp}) instead of $t^*$, the resulting profile closely matches the sharp profile but without discontinuity, figure~\ref{comparison3}.

We next present detailed results for the spherical oscillator undergoing  pulsations of both `continuous sharp' and `smooth' profiles with $a/r_0=0.35$ and $\omega=\omega_n$. The evolution of the displacement of the oscillator is reported in figure \ref{zerodamping2}(\textit{a}), confirming a quick onset of the sustained drag cancellation regime. The time-varying net force on the oscillator is shown in figures \ref{zerodamping2}(\textit{b}),(\textit{c}). Power is transferred consistently into the body during the shrinking phase of the sharp profile ($F>0$), whereas power losses are found to take place intermittently when using the smooth profile. Evolution of vortex shedding throughout a late-stage period of oscillation is depicted in figure \ref{lambda2} by visualizing the $\lambda_2$ vortex criterion of the wake for both the sharp and the smooth profiles.

Finally, we repeat the amplitude and frequency studies of the previous section using the fully coupled Navier-Stokes simulations. The results for pumping amplitude tests $0.1 \le a/r_0 \le 0.7$ at 0.1 increments and $\omega=\omega_n$ are given in figure \ref{phaseAndAmplitude}(\textit{a}). The results of 15 simulations each for the sharp and smooth profiles for pumping frequency $0.36 \le \omega/\omega_n \le 1.96$ and $0.64 \le \omega/\omega_n \le 1.7$, respectively, and $a=0.35$ are given in figure \ref{phaseAndAmplitude}(\textit{b}).

\section{Discussions}

The results presented for both the Quasi-Dynamic forcing model and the fully coupled Navier-Stokes simulations confirm the sustained cancellation of drag by added-mass pumping, figure \ref{zerodamping1} and figure \ref{zerodamping2}(\textit{a}). Pumping drives resonant growth when the excitation frequency $\omega$ is in the range  $\omega_n \sqrt{1\pm\frac{a}{2r_0}}$, figure \ref{phaseAndAmplitude}(\textit{b}), in agreement with the parametric oscillator (\ref{Omega}) for $C_{11}=\frac 12$, $m^*=1$. The limiting value of $X/a$ is approximately constant and the limit using the smooth profile is slightly greater than for the sharp profile, figure \ref{phaseAndAmplitude}(\textit{a}), in agreement with the analytic predictions (\ref{response} \& \ref{smooth_reponse}).

The deficiencies of the Quasi-Dynamic (QD) model can be assessed by comparing to the Navier-Stokes (NS) results. First, the QD model under-predicts the response for small pumping amplitude, $\frac a{r_0}<0.3$ in figure \ref{phaseAndAmplitude}(\textit{a}). The Kulegan-Carpenter number is simply $X/r_0$ for this oscillator, and it seems that for these low amplitude motions, i.e. when $X/r_0<1$, the wake history effects cannot be ignored.

Second, the QD model over-predicts the response for $\frac a{r_0}>0.4$ in figure \ref{phaseAndAmplitude}(\textit{a}). In this case the Kulegan-Carpenter number and Reynolds number are large enough that the history effects on a sphere should be negligible and the $C_D$ used should be reasonably accurate \citep{Mei1994}. Instead, the deficiency is due to the assumption of perfect energy recovery by the process of added-mass variation implicit in the effective damping coefficient $c_e$, (\ref{ce}). However, not all of the kinetic energy is recoverable in a viscous fluid and the amount depends on the shape-change rate $\sigma^*$ \citep{Weymouth2015b}. As a confirmation of this, we define a coefficient $\eta$ which scales the recovery of added-mass energy, i.e.
\begin{eqnarray}\label{ce_eta}
c_e(\eta) = \frac 12 C_D \rho A |\dot x| + \eta C_{11}\rho S \dot r 
\end{eqnarray} 
The results for $\eta=0.9$ are displayed in figure \ref{phaseAndAmplitude}(\textit{a}), and the excellent agreement with the NS results indicates that the process of added-mass recovery for this range of $\sigma^*$ is approximately 90$\%$ efficient.

Finally, we note the difference in the wake patterns shown for the two pumping profiles in figure~\ref{lambda2}, despite using the same pumping amplitude and frequency and responding with similar oscillation extent. While the same core structures exist in both results due to shedding of starting and stopping vortices, the complete cancellation of drag achieved when pumping with the sharp profile results in minimal and stable vortex structures. In contrast the smooth profile, which only achieves zero power loss on average, results in a more unstable wake pattern.

\section{Conclusions}

In this work a submerged body that varies its volume periodically is studied as a model system for sustained drag cancellation by added-mass pumping.  While the role of added-mass variation on thrust is hard to asses in general self-propelled systems, this study successfully segregates the contribution from added-mass variation and highlights its role as a source of thrust. We find that the relationship which links oscillation amplitude and extent of the pulsation is approximately linear and weakly dependent on the kinematics of the actuation. We also find that large-scale oscillation requires that the period of actuation be approximately half that of the natural period of the system, as in classic parametric resonance.

The `sharp' pumping kinematic routine which satisfies the condition for zero-damping was easily derived and has a physically intuitive form, figure \ref{comparison3}, and a striking qualitative resemblance with the propulsion routine of those organisms which are known to exploit a similar fluid dynamics phenomenon for travelling in water \citep{Trueman1968}. 

The results of this analysis shed light on the role of added-mass recovery in the context of aquatic propulsion and resonance. The implications associated with this phenomenon are important for a variety of real applications including the design of energy harvesting devices as well as the design and control of bioinspired underwater vehicles which exploit shape variations as a mean of propulsion.

\vspace{0.5cm}
\noindent
We wish to acknowledge the Lloyd's Register Foundation for the support to this work.

\bibliographystyle{jfm}

\bibliography{GabeKekko.bib}

\begin{thebibliography}{19}
\expandafter\ifx\csname natexlab\endcsname\relax\def\natexlab#1{#1}\fi

\bibitem[Candelier {\em et~al.\/}(2011)Candelier, Boyer \&
  Leroyer]{Candelier2011}
{\sc Candelier, F., Boyer, F. \& Leroyer, A.} 2011 Three-dimensional extension
  of lighthill's large-amplitude elongated-body theory of fish locomotion. {\em
  Journal of Fluid Mechanics\/} {\bf 674}, 196--226.

\bibitem[Daniel(1984)]{Daniel1984}
{\sc Daniel, T.L.} 1984 Unsteady aspects of aquatic locomotion. {\em American
  Zoologist\/} {\bf 24(1)}, 121--134.

\bibitem[Daniel(1985)]{Daniel1985}
{\sc Daniel, T.~L.} 1985 Cost of locomotion: unsteady medusan swimming. {\em
  Journal of Experimental Bilogy\/} {\bf 119}, 149--164.

\bibitem[Giorgio-Serchi {\em et~al.\/}(2016)Giorgio-Serchi, Arienti \&
  Laschi]{Giorgio-Serchi2016}
{\sc Giorgio-Serchi, F., Arienti, A. \& Laschi, C.} 2016 Underwater soft-bodied
  pulsed-jet thrusters: Actuator modeling and performance profiling. {\em
  International Journal of Robotics Research\/} .

\bibitem[Johnson {\em et~al.\/}(1972)Johnson, Soden \& Trueman]{Johnson1972}
{\sc Johnson, W., Soden, P.~D. \& Trueman, E.~R.} 1972 A study in met
  propulsion: an analysis of the motion of the squid, \textit{Loligo Vulgaris}.
  {\em Journal of Experimental Biology\/} {\bf 56}, 155--165.

\bibitem[Kanso(2009)]{Kanso2009}
{\sc Kanso, E.} 2009 Swimming due to transverse shape deformations. {\em
  Journal of Fluid Mechanics\/} {\bf 631}, 127--148.

\bibitem[Kanso \& Newton(2009)]{Kanso2009b}
{\sc Kanso, E. \& Newton, P.~K.} 2009 Passive locomotion via normal-mode
  coupling in a submerged spring-mass system. {\em Journal of Fluid
  Mechanics\/} {\bf 641}, 201--215.

\bibitem[Lavroskii \& Formal'skii(1993)]{Lavroskii1993}
{\sc Lavroskii, E.~K. \& Formal'skii, A.~M.} 1993 Optimal control of the
  pumping and damping of a swing. {\em Journal of Applied Mathematics and
  Mechanics\/} {\bf 57}, 311--320.

\bibitem[Lighthill(1960)]{Lighthill}
{\sc Lighthill, M.J.} 1960 Note on the swimming of slender fish. {\em Journal
  of Fluid Mechanics\/} {\bf 9}, 305--317.

\bibitem[Maertens \& Weymouth(2015)]{Weymouth2015}
{\sc Maertens, A.~P. \& Weymouth, G.~D.} 2015 Accurate cartesian-grid
  simulations of near-body flows at intermediate reynolds numbers. {\em Journal
  of Computational Physics\/} {\bf 283}, 106--129.

\bibitem[Mei(1994)]{Mei1994}
{\sc Mei, R.} 1994 Flow due to an oscillating sphere and an expression for
  unsteady drag on the sphere at finite reynolds number. {\em Journal of Fluid
  Mechanics\/} {\bf 270}, 133--174.

\bibitem[Morison {\em et~al.\/}(1950)Morison, O'Brien, Johnson \&
  Schaaf]{Morison1950}
{\sc Morison, J.~R., O'Brien, M.P., Johnson, J.W. \& Schaaf, S.A.} 1950 The
  force exerted by surface waves on piles. {\em Petroleum Transactions\/} {\bf
  189}, 149--154.

\bibitem[Roura \& Gonzalez(2010)]{Roura2010}
{\sc Roura, P. \& Gonzalez, J.~A.} 2010 Towards a more realistic description of
  swing pumping due to the exchange of angular momentum. {\em European Journal
  of Physics\/} {\bf 31}, 1195--1207.

\bibitem[Saffman(1967)]{Saffman}
{\sc Saffman, P.~G.} 1967 The self-propulsion of a deformable body in a perfect
  fluid. {\em Journal of Fluid Mechanics\/} {\bf 28}, 385--389.

\bibitem[Spagnolie \& Shelley(2009)]{Shelly2009}
{\sc Spagnolie, S.~E. \& Shelley, M.~J.} 2009 Shape-changing bodies in fluids:
  hovering, ratcheting and bursting. {\em Physics of Fluids\/} {\bf 21}, 1--13.

\bibitem[Trueman \& Packard(1968)]{Trueman1968}
{\sc Trueman, E.~R. \& Packard, A.} 1968 Motor performances of some
  cephalopods. {\em Journal of Experimental Biology\/} {\bf 49}, 495--507.

\bibitem[Weymouth {\em et~al.\/}(2015)Weymouth, Subramaniam \&
  Triantafyllou]{Weymouth2015b}
{\sc Weymouth, G.D., Subramaniam, V. \& Triantafyllou, M.S.} 2015 Ultra-fast
  escape maneuver of an octopus-inspired robot. {\em Bioinspiration $\&$
  Biomimetics\/} {\bf 10}, 1--7.

\bibitem[Weymouth \& Triantafyllou(2013)]{Weymouth2013}
{\sc Weymouth, G. \& Triantafyllou, M.S.} 2013 Ultra-fast escape of a
  deformable jet-propelled body. {\em Journal of Fluid Mechanics\/} {\bf 721},
  367--385.

\bibitem[Weymouth \& Yue(2011)]{Weymouth2011}
{\sc Weymouth, G.~D. \& Yue, D. K.-P.} 2011 Boundary data immersion method for
  cartesian-grid simulations of fluid-body interaction problems. {\em Journal
  of Computational Physics\/} {\bf 230}, 6233--6247.

\end{thebibliography}

\end{document}